\documentclass[useAMS,usenatbib,usegraphicx]{mn2e}

\usepackage{definitions}
\usepackage{rotating}
\usepackage{amsfonts}
\usepackage{graphicx}
\usepackage{amsbsy}
\renewcommand{\v}[1]{\textbf{\textit{#1}}}

\let\mr=\mathrm

\newcommand*{\dd}{\, \mathrm{d}}

\newcommand{\bq}{\begin{equation}}
\newcommand{\eq}{\end{equation}}

\newcommand{\vtheta}{\boldsymbol{\theta}}

\newcommand{\LCDM}{\Lambda\mathrm{CDM}}

\def\bigexpec#1{\left\langle#1\right\rangle}

\title[Model selection for dark energy]{Bayesian model selection for dark energy using weak lensing forecasts}
\author[I. Debono]{Ivan~Debono\thanks{E-mail: ivan.debono@apc.univ-paris7.fr}\\
{APC, AstroParticule et Cosmologie, Universit\'{e} Paris Diderot, CNRS/IN2P3, CEA/lrfu, Observatoire de Paris, Sorbonne Paris Cit\'{e}}\\{10, rue Alice Domon et L\'{e}onie Duquet, 75205 Paris Cedex
13, France.} \\{LESIA, Observatoire de Paris, CNRS, UPMC, Universit\'{e} Paris Diderot,}\\{5 place Jules Janssen, 92195 Meudon, France.}}
\begin{document}

%\date{Accepted 1988 December 15. Received 1988 December 14; in original form 1988 October 11}

\pagerange{\pageref{firstpage}--\pageref{lastpage}} \pubyear{2013}

\maketitle
\label{firstpage}

\begin{abstract}
The next generation of weak lensing probes can place strong constraints on cosmological parameters by measuring the mass distribution and geometry of the low redshift universe. We show that a future all-sky tomographic cosmic shear survey with design properties similar to \textit{Euclid} can provide the statistical accuracy required to distinguish between different dark energy models. Using a fiducial cosmological model which includes cold dark matter, baryons, massive neutrinos (hot dark matter), a running primordial spectral index and possible spatial curvature as well as dark energy perturbations, we calculate Fisher matrix forecasts for different dynamical dark energy models. Using a Bayesian evidence calculation we show how well a future weak lensing survey could do in distinguishing between a cosmological constant and dynamical dark energy. 
\end{abstract}


\section{Introduction}
\label{Intro}

The science of cosmology finds itself at a critical point where it has to make sense of the large quantity of data that has become available. Data from various astrophysical sources (large scale structure \citep[][SDSS]{Ahn:2013aa}; the cosmic microwave background \citep[][\textit{Planck}]{Planck-Collaboration:2013aa}; supernovae (\citealt[][SCP]{Goldhaber:2009aa}); weak lensing \citep{Schrabback:2009}) have allowed us to measure the parameters in our cosmological model with ever increasing precision. 

The $\LCDM$ Concordance Model has been very successful at explaining a host of observations with only six parameters. In this concordance cosmology, initial quantum fluctuations are believed to have seeded perturbations in the matter distribution, leading to the Large Scale Structure we observe today. Within this model, the Universe is composed only of a small proportion of baryons (about 5 per cent), the rest being dark matter (about 25 per cent, which can be hot or cold) and dark energy. The success of the Concordance Model has been its ability to include physical effects at different scales, from primordial nucleosynthesis to Large Scale Structure evolution, in one coherent theory. Yet there still remains the problem of parameter accuracy. Given the data, we can measure the parameters within a theoretical model to a given precision, but is the model itself correct? In other words, together with the problem of parameter estimation, there is the problem of model selection.

The statistical questions facing cosmologists pose some particular problems. We observe a finite region of our Universe, which is itself a single realisation of the cosmological theory. In other words, we have a single data point for the cosmological model. We need to make decisions based on this incomplete information. Bayesian inference provides a quantitative framework for plausible conclusions (see \citet{Hobson:2010aa} for a detailed presentation). We can identify three levels of Bayesian inference:
\begin{enumerate}
\item Parameter inference (estimation): we assume that a model $M$ is true, and we select a prior for the parameters $P(\vtheta | M)$.
\item Model comparison: there are several possible models $M_i$. We find the relative plausibility of each in the light of the data $D$.
\item Model averaging: there is no clear evidence for a best model. We find the inference on the parameters which accounts for the model uncertainty.
\end{enumerate}

At the first level of Bayesian inference, we can estimate the allowed parameter values of a model. Next, we can ask which parameters we should include in our cosmological model. Although current data are consistent with the six-parameter $\Lambda$CDM model, there are more than twenty candidate parameters which might be required by future data \citep[see][]{Liddle:2004aa}. We cannot simply include all possible parameters to fit the data, since each one will give rise to degeneracies that weaken constraints on other parameters, including the $\Lambda$CDM parameter set \citep[see e.g.][]{Debono:2009}. 

The goal in data analysis is usually to decide which parameters need to be included in order to explain the data. For cosmologists, those extra parameters must be physically motivated. That is, we need to know the physical effects to which our data are sensitive, so that we can relate these effects to physics. At the current state of knowledge, we have to acknowledge the possibility of more than one model. We therefore require a consistent method to discard or include parameters. This is the second level of Bayesian inference -- model selection.  

It is only recently that cosmologists have focussed on model selection \citep[see, e.g.][]{Jaffe:1996, Mukherjee:2006, Kilbinger:2009a, Wraith:2009}, when the astrophysical data began to have the necessary statistical power to enable model testing. With the next generation of astrophysical probes in the pipeline, model selection is likely to grow in importance \citep[for a comprehensive review, see][and note figure 1 therein]{Trotta:2008aa}.

Bayesian model selection cannot be completely free of assumptions. In cosmology, there is some model structure which depends on a number of unverifiable hypotheses about the nature of the Universe. This important point is addressed by \citet{Ellis:1975aa,Ellis:2009,Ellis:2006aa}, who identifies the Copernican Principle as one such hypothesis.  It is worth stating the assumptions that we use throughout this paper:
\begin{enumerate}
\item The Universe is isotropic.
\item The Universe is homogeneous (beyond a certain scale). 
\item Gravitational interactions are described by General Relativity.
\end{enumerate}
These assumptions mean that the Universe can be described by General Relativity and a Friedmann-Robertson-Walker cosmology.
 
There is now firm observational evidence suggesting that our universe entered a recent stage of accelerated expansion. The physical mechanism behind this expansion rate remains unclear and there exists a wealth of potential models. In the framework of General Relativity applied to a homogeneous and isotropic universe, the acceleration could be produced by a new isotropic comoving perfect fluid with negative pressure, called dark energy \citep[and references therein]{Polarski:2010aa, Polarski:2013aa}. Models explaining the present accelerated expansion are called dark energy models. The simplest dark energy candidate is a fluid whose pressure satisfies $p_\Lambda = - \rho_\Lambda$ or $w_\Lambda \equiv {p_\Lambda }/{\rho_\Lambda}=-1$. This is equivalent to a positive cosmological constant $\Lambda$ added to the Einstein equations. The real question for physicists lies here. Are we in the presence of a new component of the universe or is this merely an additional term in the equations describing gravitational interactions? If the former, what is the nature of this component? If the latter, why the additional constant? 

In order to produce an accelerated expansion at the present epoch, the equation of state parameter should satisfy the conservative bound $w_\mr{DE}={p_\mr{DE} }/{\rho_\mr{DE}}< -0.5$. Observations suggest a lower value, close to $-1$. Is the value constant, or does it vary with time?  In the cosmological context, the question in Bayesian inference terms is whether there is evidence that we need to expand our cosmological model beyond $\Lambda$CDM to fit this data \citep[see e.g.][]{March:2011aa}.

One of the fundamental problems in modern cosmology is to understand this dark energy component, which constitutes around 70 per cent of the Universe's energy density \citep{DETF, Peacock:2006}.  To distinguish between theories the determination of the dark energy equation of state $w$ is essential since some models can result in very different expansion histories. The next generation of cosmic shear surveys show exceptional potential for constraining the dark energy equation of state $w(z)$  \citep{DETF, Peacock:2006} and have the advantage of directly tracing the dark matter distribution (see \citealt{Hoekstra:2008} for a review).

Evidence for a departure from $\Lambda$CDM may come from any sector. It is therefore pertinent to examine extensions of this model in all sectors of interest within the Bayesian framework. 

The first extension to $\Lambda$CDM is the addition of neutrino parameters. Massive neutrinos have a non-negligible effect on the matter power spectrum, especially the non-linear part \citep{Elgaroy:2005,Hannestad:2006,Ichiki:2008,Kitching:2008aa,Lahav:2009,de-Bernardis:2009aa,Debono:2009,Melchiorri:2012aa,Vanderveld:2013aa}. The level of precision of future weak lensing surveys is such that the lensing signal is sensitive to the contribution of the non-linear part of the matter power spectrum, where neutrino physics plays a significant role. This requires the inclusion of neutrino parameters in our calculations. 

In the primordial power spectrum sector, the recent CMB anisotropy observations from the \textit{Planck} probe rule out exact scale invariance at over $5 \sigma$ \citep{Planck-Collaboration:2013ab}. Although there is no statistically significant evidence for scale dependence, the data suggests caution in employing an overly simple parametrization. For this reason, we allow for possible departures from a scale-invariant primordial power spectrum.

Cosmic shear surveys have the potential to constrain all sectors of our cosmological model.  As shear measurements depend on the initial seeds of structure, weak lensing can be used to constrain initial power spectrum parameters \citep[see e.g.][]{Liu:2009} which are central to our understanding of the inflationary model \citep[see e.g.][]{Hamann:2007}. Shear measurements have also been used to complement neutrino constraints from particle physics  (\citealt{Tereno:2008}; \citealt*{Ichiki:2008}) and galaxy surveys \citep*{Takada:2006}, and future weak lensing surveys will provide bounds on the sum of the neutrino masses, the number of massive neutrinos and the hierachy (\citealt*{Hannestad:2006, Kit2008b}; \citealt{de-Bernardis:2009aa}; \citealt{Hamann:2012aa}; \citealt{Audren:2013aa}).
 
In this work we will consider weak cosmic shear from a future all-sky survey similar to the European Space Agency mission \textit{Euclid} \citep[see][]{Amendola:2012aa}. The main scientific objective of \textit{Euclid} is to understand the origin of the accelerated expansion of the Universe by probing the nature of dark energy. It could potentially test for departures from the current standard models \citep[see e.g.][]{Heavens:2007,Zhao:2012aa,Jain:2010aa}. The main goal is therefore to test for departures from $\Lambda$CDM. 

In terms of model selection, we can identify four hypotheses about the equation of state parameter $w(a)$ which form the `hypothesis space' of the dark energy paradigm (see Section \ref{DE_section} for notation):
\begin{description}
\item $H_0$: Cosmological constant  $\Lambda$: $w(a) = -1$
\item $H_1$: Constant, but non-$\Lambda$: $w(a) = w_0$
\item $H_2$: Simple evolving: $w(a) = w_0 +(1-a)w_a$
\item $H_3$: Some more complex form of $w(a)$
\end{description}
At present, there is no evidence which justifies assigning a higher probability to a particular model. This is a statement on our prior knowledge, which is based on the accumulation of information from a multitude of experiments \citep[see][]{Brewer:2009aa}. In this paper, we justify assigning equal probabilities to each model because we are testing two at a time. We will consider the first three models in the above list.  

This paper is organized as follows. In Section \ref{Bayesian_inference}, we describe the Bayesian framework and our model evidence method. Our cosmological and weak lensing formalism is described in Section \ref{Method}. Parameter estimation and model selection results are presented in sections \ref{Parameter_estimation} and \ref{Model_selection}.

\section{Bayesian inference}
\label{Bayesian_inference}

The formalism of Bayesian inference is based on Bayes' theorem, which is derived from the product rule in probability theory:
\bq\label{eqn1}
p(\vtheta | D,M)=\frac{p(D | \vtheta,M)p (\vtheta | M)}{p( D | M)},
\eq
where the left-hand side is the posterior probability for the (in general) multi-dimensional vector of unknown model parameters $\vtheta$ of length $n$ given the data $D$ under model $M$, and $p (D | M)$ is the Bayesian evidence or model likelihood. The latter is the probability of observing the data $D$ given that the model $M$ is correct. 

Model selection or testing usually involves the calculation of the evidence. This may be expressed as the multi-dimensional integral of the likelihood over the prior
\bq
p (D | M)=\int L \dd D = \int  p(D | \vtheta,M) p(\vtheta | M) \dd\vtheta ,
\eq
where $L=L(\vtheta)$ is the likelihood function and $\dd D =p(\vtheta)\dd \vtheta$ is the element of prior mass. 

Here we are interested in model selection from among two models, i.e. the probability that the model $M_0$ is correct instead of an alternative model $M_1$. This is given by the posterior relative probabilities of the two models:
\bq
\frac{p(M_0|D)}{p(M_1|D)}=\frac{p(M_0)}{p(M_1)}\frac{\int p(D|\vtheta_0,M_0)p(\vtheta_0|M_0) \dd\vtheta_0}{\int p(D|\vtheta_1,M_1)p(\vtheta_1|M_1) \dd\vtheta_1}.
\eq
 
If we assume that the models have the same probability of being correct (that is, if we have noncommittal priors), we have \bq \frac{p(M_0)}{p(M_1)}=1.\eq The ratio then simplifies to give us the Bayes factor, or the ratio of the probabilities that model 0 (null hypothesis) is correct over model 1 given the data:
\bq B_{01}\equiv \frac{\int  p(D | \vtheta_0,M_0) p(\vtheta_0 | M_0) \dd \vtheta_0}{\int  p(D | \vtheta_1,M_1) p(\vtheta_1 | M_1) \dd \vtheta_1} .\eq

The Bayes factor is then the ratio of the evidence for two competing models $M_0$ and $M_1$, or the ratio of posterior odds:
\bq
B_{01}\equiv\frac{p(D | M_0)} {p(D | M_1)}=\frac{p(M_0|D)}{p(M_1|D)}.
\eq

When we have two models, and we convert from probabilities to odds, the Bayes factor acts like an operator on prior beliefs:
\bq \frac{p(M_0 | D)}{p(M_1 | D)} = B_{01} \times  \frac{p(M_0)}{p(M_1)}
\eq

\bq \textrm{Posterior odds} = \textrm{Bayes factor} \times \textrm{Prior odds}
\eq
 
We assume separable priors in each parameter over the range $\Delta \theta$. This is usually valid in many cosmological applications. Hence 
\bq p(\vtheta | M)=(\Delta\theta_1\cdots\Delta\theta_n)^{-1}.
\eq
The width of the prior range may influence the Bayes factor. While the prior range should be large enough to contain most of the likelihood volume, an arbitrarily large prior can result in an arbitrarily small evidence. For a discussion on the dependence of evidence on the choice of prior see e.g. \citet{Trotta:2007} and \citet{Brewer:2009aa}.

The Bayes factor becomes

\bq 
B_{01}=\frac{\int p(D|\vtheta_0,M_0) \dd\vtheta_0}{\int p(D|\vtheta_1,M_1) \dd\vtheta_1}\frac {(\Delta\theta_1\cdots\Delta\theta_m)_1}{(\Delta\theta_1\cdots\Delta\theta_n)_0}
\label{Bayes_f} 
\eq 
where $n$ and $m$ are the number of parameters in models $M_0$ and $M_1$, respectively. If $M_0$ is the simpler model, with its $n$ parameters common to $M_1$, which has $p=m-n$ extra free parameters, then $M_0$ may be considered as a special case of a more general model $M_1$. For such nested models, assuming separable priors, the ratio of prior hypervolumes simplifies to
\bq 
\frac {(\Delta\theta_1\cdots\Delta\theta_m)_1}{(\Delta\theta_1\cdots\Delta\theta_n)_0}=\Delta\theta_{n+1}\cdots\Delta\theta_{n+p}. 
\eq
Then the Bayes factor can be written as
\textbf{\bq
B_{01}=\frac{p(\theta_{n+1}\cdots\theta_{n+p}|D,M_1)}{p(\theta_{n+1}\cdots\theta_{n+p}|M_1)}\bigg|_{\theta_{n+1}\cdots\theta_{n+p}=0}.
\eq}
This is the Savage-Dickey density ratio or SDDR \citep{Dickey1971}.

In this paper, we will calculate the SDDR from the Fisher matrix using the method described in \citet{Lazarides:2004aa} and \citet{Heavens:2007}. Here we follow the notation used in the latter work.

For a future experiment, the data $D$ do not exist, so we compute the expectation value of the Bayes factor, given the statistical properties of $D$ (which we calculate from the survey configuration). We also make two further approximations:
\begin{enumerate}
\item Since $B$ is a ratio, we approximate $\langle B \rangle$ by the ratio of expected values (rather than the expected value of the ratio). This holds if the evidences are sharply peaked. 
\item We use the Laplace approximation, where we assume that the likelihoods are well-described by a multivariate Gaussian. 
\end{enumerate}
The second approximation is used in the Fisher matrix method in parameter estimation \citep{Fisher1935}, which gives the lower bound on the accuracy with which we can estimate model parameters from a given data set \citep[see][]{Tegmark:1997}. This assumes that the second-order Taylor expansion of $\ln L$ is valid over a sufficiently wide region in parameter space around the maximum. For a \textit{Euclid}-like weak lensing survey, \citet{Audren:2013aa} find that the expected likelihoods for the model parameters of a $\Lambda$CDM model including massive neutrinos are very close to a multivariate Gaussian. 

The Fisher matrix, defined as 
\bq
F_{\alpha\beta}=-\bigexpec{\frac{\partial^2\ln L}{\partial\theta_\alpha \partial\theta_\beta } },
\eq
where $\theta_\alpha$ is the vector of model parameters, is the expectation value of the covariance matrix near the maximum-likelihood point. The Fisher matrix approach allows us to forecast the errors around the parameter values, and can therefore be used to forecast the precision with which a given experiment will estimate model parameters. 

Using the expression for the Fisher matrix, the expected likelihoods for the model $M$ parameters are then
\bq
\langle p(D|\vtheta,M)\rangle=L^\ast \exp \left(-\frac{1}{2}(\theta_\alpha-\theta_\alpha^\ast)F_{\alpha\beta}(\theta_\beta-\theta_\beta^\ast)\right),
\eq where $\ast$ indicates the peak value. 

If we assume that the posterior probability densities are small at the boundaries of the prior volume, the integration over the multivariate Gaussian in equation (\ref{Bayes_f}) becomes $(2\pi)^{n/2}(\det F)^{1/2}$, where $n$ is the number of parameters in model $M$.

For nested models, the expected value of the Bayes factor is then:
\bq
\langle B_{01}\rangle=(2\pi)^{-p/2}\frac{\sqrt{\det F_1}}{\sqrt{\det F_0}}\frac{{L^\ast}_0}{{L^\ast}_1}\prod_{q=1}^p\Delta\theta_{n'+q}.
\eq
At the maximum of the expected likelihood in the extended model $M_0$, the parameters $\alpha$ shift by $\delta\alpha$ from their values in the simpler model $M_1$ to compensate for the fact that the additional $p$ parameters differ by $\delta\psi$ from zero. This shift is given by:
\bq
\delta\theta_\alpha=-(F_1^{-1})_{\alpha\beta} G_{\beta_\zeta}\delta\psi_\zeta \quad ; \alpha,\beta=1\ldots n,\quad\zeta=1\ldots p
\eq
where $G$ is an $n$ by $p$ block of the Fisher matrix $F_0$. The ratio of the likelihoods calculated from the Fisher matrix is then
\bq \frac{{L^\ast}_0} {{L^\ast}_1} = \exp\left(-\frac{1}{2}\delta\theta_\alpha [{F}_1]_{\alpha\beta} \delta\theta_\beta \right)
\eq
 We therefore obtain the final expression for the expected value of the Bayes factor as a special case of the SDDR:
\[
\langle B_{01}\rangle=(2\pi)^{-{p}/{2}}\frac{\sqrt{\det F_1}}{\sqrt{\det F_0}}\exp\left(-\frac{1}{2}\delta\theta_\alpha [{F}_1]_{\alpha\beta} \delta\theta_\beta \right)\prod_{q=1}^p\Delta\theta_{n+q}.
\]
\bq \eq
This allows us to evaluate the expected evidence without having to calculate the likelihood at all interesting points in parameter space, making the calculation less computationally demanding. 

\citet{Lazarides:2004aa} give a similar expression for the SDDR which explicitly illustrates the underlying concepts:
\bq \ln B_{01}=\mathcal{L}_{01}+\mathcal{C}_{01}+\mathcal{F}_{01},\eq where $\mathcal{L}$ is likelihood, $\mathcal{C}$ is the posterior volume and $\mathcal{F}$ is the prior structure.  The model evidence thus incorporates a trade-off between modelling the data (i.e. goodness of fit) and remaining consistent with our prior (i.e. simplicity or negative complexity). The latter, through the posterior volume and prior structure terms, can be interpreted as Occam's razor, or the principle of parsimony in scientific theories.

A Bayes factor (in contrast to a likelihood ratio) thus says which of two competing models is better at providing a simple yet accurate explanation of the data (see \citet{March:2011aa} for an interesting discussion in relation to $\Lambda$CDM).

Evidence values need to be interpreted in order to make a decision on competing models. In this study we will use the convention summarised in Table \ref{Jeffreys}. Note that different authors assign different descriptions to evidence values (e.g. \citet{Jeffreys:1961}, \citet{Trotta:2007aa}, \citet{Kass:1995}). One should bear in mind that the interpretation of Bayesian evidence depends on the problem in question \citep[see e.g.][]{Efron:2001}. In Table \ref{Jeffreys} we list the terminology given by \citet{Jeffreys:1961} and \citet{Trotta:2007aa}, including the translation of evidence values to model probabilities. 

\begin{table}
\caption{Jeffrey's scale for the strength of evidence when comparing two models $M_0$ against $M_1$. The probability is the posterior probability of the favoured model, assuming non-committal priors on the two models, and assuming that the two models fill all the model space. The threshold values are set empirically, and are roughly based on the standard interpretation of betting odds. }
\begin{center}
\begin{tabular}{@{} lllll @{}}
\hline
$| \ln B |$ 		& Odds 		& Probability 	& \multicolumn{2}{c}{Evidence}\\
&&&\citeauthor{Jeffreys:1961}&\citeauthor{Trotta:2007}\\
\hline
$<1.0$	 & $\la 3:1$ 	& $<0.750$ 	& Inconclusive & Inconclusive\\
$1.0$ 	& $\sim 3:1$ 		& $0.750$ &Substantial	& Positive \\
$2.5$ 	& $\sim12:1$ 		& $0.923$ &Strong	&  Moderate \\
$5.0$ 	& $\sim150:1$ 		& $0.993$ &Decisive	 &  Strong \\
\hline
\end{tabular}
\end{center}
\label{Jeffreys}
\end{table}

Inconclusive evidence for one model means that the alternative model cannot be distinguished from the null hypothesis. This occurs when $|\ln B_{01}|<1$. A positive Bayes factor $\ln B_{01}>1$ favours model $M_0$ over $M_1$ with odds of $B_{01}$ against 1. In this respect, Bayesian model selection is different from frequentist goodness-of-fit tests, since we cannot reject a hypothesis unless an alternative hypothesis is available that fits the facts better. If we take the example of dark energy models, it means that a claim such as `$\Lambda$CDM is false' is not useful, and we need a more specific model in our set of testable hypotheses \citep[see][]{Trotta:2008aa}.

\section{Method}
\label{Method}

\subsection{Cosmology}

We start by describing our simplest parameter set. Our 9-parameter FRW cosmological model contains baryonic matter, cold dark matter (CDM) and dark energy, to which we add massive neutrinos (i.e. hot dark matter -- HDM). 

We allow for a non-flat spatial geometry by including a dark energy density parameter $\Omega_\mr{DE}$ together with the total matter density $\Omega_m$, such that in general $\Omega_m+\Omega_\mr{DE}\neq 1$. 

Our 9-parameter space consists of:
\begin{enumerate}
\item Total matter density -- $\Omega_m$ (which includes baryonic matter, HDM and CDM) 
\item Baryonic matter density -- $\Omega_b$
\item Neutrinos (HDM) -- $m_\nu$ (total mass in eV), $N_\nu$ (number of  massive neutrino species)
\item Dark energy density -- $\Omega_{\mathrm{DE}}$
\item Hubble parameter -- $h$ in units of $100\,\mr{km}^{-1}\mr{Mpc}^{-1}$
\item Primordial power spectrum parameters -- $\sigma_8$ (amplitude), $n_s$ (scalar spectral index), $\alpha$ (its running)
\end{enumerate}
For simplicity, we shall refer to this fiducial cosmology as $\Lambda$CDM. We choose fiducial parameter values based on the \textit{Planck} 2013 results \citep{Planck-Collaboration:2013aa}, with the exception of the scalar spectral index, which we set to 1 (\textit{Planck} data indicate a slightly lower value which is incompatible with 1 \citep[see][]{Planck-Collaboration:2013aa,Planck-Collaboration:2013ab}). Our fiducial values are: $\Omega_m$=0.31, $\Omega_b=0.048$, $m_\nu=0.25\,\mr{eV}$, $N_\nu=3$, $h=0.67(100\,\mr{km}^{-1}\mr{Mpc}^{-1})$, $\Omega_{\mr{DE}}=0.69$, $n_s=1$, $\alpha=0$, $\sigma_8=0.82$. The reionization optical depth, used in the power spectrum calculations but not included in the Fisher matrix, is set to $\tau=0.09$.

\subsection{Matter power spectrum}
\label{Power_spectrum}

The matter power spectrum is defined as:
\bq 
 \langle \delta(\mathbf{k})\delta^\ast(\mathbf{k}')\rangle= {(2\pi)}^3\delta_D^3(\mathbf{k}-\mathbf{k}') P(k)
 \eq 
 and can be modelled by: \bq P(k,z)=\frac{2\pi^2}{k^3}A_sk^{n_s(k)+3}{T^2(k,z)}\left(\frac{D(z)}{D(0)}\right)^2,\eq where $A_s$ is the normalisation parameter, $T(k,z)$ is the transfer function and $D(z)$ is the growth function. The primordial spectral index is denoted by $n_s(k)$. 
 
In our cosmological model, the shape of the primordial power spectrum is of particular interest, since it may mimic some of the small-scale power damping effect of massive neutrinos. In the concordance model, the primordial power spectrum is generally parametrized by a power-law \citep[see e.g.][]{Kosowsky:1995, Bridle:2003} \bq\mathcal{P}_\chi(k)=A_s\left(\frac{k}{k_{0}}\right)^{n_s-1},\eq where $A_s$ is the power spectrum amplitude used as a normalisation parameter in the previous equation.  We parametrize the running of the spectral index by using a second-order Taylor expansion of $\mathcal{P}_\chi$ in log-log space, defining the running as $\alpha = \mr{d} n_s/\mr{d}\ln k |_{k_0}$, so that the primordial power spectrum is scale-dependent, with the scalar spectral index defined by \citep{Spergel:2003, Hannestad:2002} 
\bq n_s(k)=n_s(k_0)+\frac{1}{2}\frac{ \mr{d} n_s}{ \mr{d} \ln k}\bigg|_{k_0} \ln\left(\frac{k}{k_0}\right),\eq where $k_0$ is the pivot scale. This parametrization is motivated by standard slow-roll inflation theory. We use a fiducial value of ${k_0=0.05 \mr{Mpc^{-1}}}$ for the primordial power spectrum pivot scale.

Our fiducial cosmological model assumes a total of three neutrino species (i.e. $N_\mr{massless}+N_\nu=3$), with degenerate masses for the most massive eigenstates, i.e. if $m_\nu$ is the total neutrino mass, then 
\bq m_\nu=\sum^{N_\nu}_{i=0}m_i=N_\nu m_i, \eq 
where $m_i$ is the same for all eigenstates. The temperature of the relativistic neutrinos is assumed to be equal to $(4/11)^{1/3}$ of the photon temperature \citep{KolbTurner1990}. We model $N_\nu$, the number of massive (non-relativistic) neutrino species, by a continuous variable. 

Neutrino oscillation experiments do not, at present, determine absolute neutrino mass scales, since they only measure the difference in the squares of the masses between neutrino mass eigenstates \citep{Quigg:2008}. Cosmological observations, on the other hand, can constrain the neutrino mass fraction, and can distinguish between different mass hierarchies (see e.g. \citealt{Elgaroy:2005}, \citealt{Jimenez:2010aa}, \citealt{Abazajian:2011aa}, \citealt{Jimenez:2013aa} for a review of the methods). In particular, Bayesian model selection based on the SDDR using weak lensing has been applied to neutrino parameters by \citet{Kitching:2008aa} and \citet{de-Bernardis:2009aa}.

We use the numerical Boltzmann code \textsc{CAMB} \citep{CAMB} to calculate the linear matter power spectrum. This includes the contribution of baryonic matter, cold dark matter, dark energy and massive neutrino oscillations. We use the \citet{Smith:2003aa} \textsc{halofit} fitting formula to calculate the non-linear power spectrum, with the modification suggested by \citet{Bird2012}, which models the induced neutrino suppression over our power spectrum scales and redshifts better than \textsc{halofit}. The power spectrum is normalised using $\sigma_8$, the root mean square amplitude of the density contrast inside an $8\,h^{-1}\mr{Mpc}$ sphere. 

\subsection{Dark energy parametrization}
\label{DE_section}

This paper examines the following question: Is dark energy $\Lambda$? How well will the future \textit{Euclid} probe be able to answer this?

We therefore extend our fiducial $\Lambda$CDM parameter space by adding two dark energy parameters: the equation of state parameter at the present epoch $w_0$ and its variation $w_a$. The dynamical dark energy equation of state parameter, $w=p/\rho$, is expressed as function of redshift and is parametrized by  a first-order Taylor expansion in the scale factor $a$ \citep{ChevPol2001, Linder:2003}: \bq w(a)=w_0 + (1 -a)w_a,\eq where $a=(1+z)^{-1}$. This parametrization is motivated by the quintessence model, in which dark energy is some minimally coupled scalar field, slowly rolling down its potential such that it can have negative pressure. Scalar field models typically have a time-varying $w\geqslant -1$. The model selection question on $\Lambda$CDM can therefore be reformulated as follows: Is $(w_0,w_a)=(-1,0)$?

Dark energy affects the matter power spectrum in three ways. Firstly, its density $\Omega_\mr{DE}$ changes the normalisation and $k_{eq}$, the point at which the power spectrum turns over. Secondly, $\Omega_\mr{DE}$ and the dark energy equation of state parameter $w$ change the growth factor at late times by changing the Hubble rate. In addition to this, for departures from a cosmological constant the shape of the matter power spectrum on large scales is affected through dark energy perturbations. 

Testing for the presence of any small deviations from $w=-1$ using future probes requires a model of dark energy that allows the equation of state to evolve across this boundary multiple times. The effect of these perturbations tends to be small close to $w=-1$ \citep[see e.g.][]{Hu:2005aa,Vikman:2005aa,Caldwell:2005aa}. However, our aim is to explore as fully as possible the $(w_0-w_a)$ parameter space, including regions where $w<-1$ (the `phantom domain') and models in which $w$ is time-dependent. A single-fluid model is not self-consistent when the equation of state is allowed to cross below $-1$. Using a frequentist combination of CMB and SNIa data, \citet{Yeche:2006aa} find that the introduction of dark energy perturbations for $w>-1$ can change the position of the maximum likelihood in the $(w_0-w_a)$ plane by as much as $2\sigma$.

We therefore include dark energy perturbations in all our calculations by using the parametrized post-Friedmann (PPF) framework \citep{Hu:2007aa,Hu:2008aa} as implemented in \textsc{CAMB} \citep{Fang:2008aa,Fang:2008ab}. This allows a time-dependent equation of state $w(a)$ to cross the phantom divide multiple times.

\subsection{Weak lensing}

The cosmological probe considered in this paper is tomographic weak lensing, from which we derive our observable: cosmic shear. We base our calculations on the configuration and properties of a future all-sky tomographic weak lensing survey such as \textit{Euclid} \citep{Amendola:2012aa,Laureijs:2011aa}. The calculation of the weak lensing power spectrum closely follows the method described in \citet{AR2007} and \citet{Debono:2009}. 

In cosmic shear surveys, the observable is the convergence power spectrum. In our analysis, we calculate this quantity from the matter power spectrum via the lensing efficiency function. Our convergence power spectrum therefore depends on the survey geometry and on the matter power spectrum. We use the power spectrum tomography formalism by \citet{Hu:2004}, with the background lensed galaxies divided into 10 redshift bins. Cosmological models are then constrained by the power spectrum corresponding to the cross-correlations of shears within and between bins. The 3D power spectrum is projected onto a 2D lensing correlation function using the \citet{Limber:1953} equation:
\bq C_\ell^{ij}=\int \mr{d}z \frac{H}{D^2_A} W_i(z)W_j(z)P(k=\ell/D_A,z),\eq
where $i$, $j$ denote different redshift bins. The weighting function $W_i(z)$ is defined by the lensing efficiency: 
 \bq  
W_i(z)=\frac{3}{2}\Omega_m \frac{H_0}{H}\frac{H_0 D_{OL}}{a}\int_z^\infty \mr{d} z^{\prime}\frac{D_{LS}}{D_{OS}}P(z^{\prime}),
\eq where the angular diameter distance to the lens is $D_{OL}$, the distance to the source is $D_{OS}$, and the distance between the source and the lens is $D_{LS}$ (see \citealt{Hu:2004} for details). Our multipole range is $10<\ell<5000$. This is a good compromise between the gain in information by the inclusion of non-linear modes, and the uncertainties in the matter power spectrum calculation for sub-arcmin scales \citep{AR2007,Debono:2009}.

The galaxies are assumed to be distributed according to the following probability distribution function \citep*{Smail1994}:
\bq P(z)=z^a \exp\left[-\left(\frac{z}{z_0}\right)^b\right],\eq where $a=2$ and $b=1.5$, and $z_0$ is determined by the median redshift of the survey $z_m$ \citep[see e.g.][]{AR2007}. 

Our survey geometry follows the parameters for two configurations of an all-sky survey of the \textit{Euclid} type, with the survey area $A_s$ ranging from $15 000$ to $20 000\,\mr{sq\; degrees}$. The first configuration uses the properties defined by the `requirements', and the second uses the `goals', as defined in the \textit{Euclid} Definition Study Report \citep{Laureijs:2011aa}.  The survey parameters are shown in Table \ref{survey_params}. The median redshift of the density distribution of  galaxies is $z_\mr{median}$ and the observed number density of galaxies is $n_g$. We include photometric redshift errors $\sigma_z(z)$ and intrinsic noise in the observed ellipticity of galaxies $\sigma_\epsilon$. We follow the definition $\sigma_\gamma^2=\sigma_\epsilon^2$, where $\sigma_\gamma$ is the variance in the shear per galaxy \citep[see][]{Bartelmann:2001}.

\begin{table}
\begin{center}
\caption{Fiducial parameters for the \textit{Euclid}-type all-sky weak lensing survey considered in this paper.}
\label{survey_params}
\begin{tabular}{@{}lrr@{}}
\hline
Survey property & Requirements & Goals\\
\hline
$A_s$/sq degree & 15 000 &20 000\\
$z_\mr{median}$&0.9 & 0.9\\
$n_g/\mr{arcmin}^{2}$&30 & 40\\
$\sigma_z(z)/(1+z)$& 0.05 &0.03\\
$\sigma_\epsilon$&0.25 & 0.25\\
\hline
\end{tabular}
\end{center}
\end{table}

The Fisher matrix for the shear power spectrum is given by \citep{Hu:2004}
\bq F_{\alpha\beta} = f_\mr{sky}\sum_\ell{\frac{(2\ell+1)\Delta \ell}{2}}\mr{Tr}\left[D_{\ell\alpha}\widetilde{C}_\ell^{-1}D_{\ell\beta}\widetilde{C}_\ell^{-1}\right],\eq
where the sum is over bands of multipole $\ell$ of width $\Delta \ell$, $\mr{Tr}$ is the trace, and $f_\mr{sky}$ is the fraction of sky covered by the survey. We assume the likelihood to have a Gaussian distribution, with zero mean.  The observed power spectra for each pair $i,j$ of redshift bins are written as the sum of the lensing and noise spectra:
\bq \widetilde{C}_\ell^{ij}=C_\ell^{ij}+N_\ell^{ij}.\eq
The derivative matrices are given by
\bq [D_{\alpha}]^{ij}=\frac{\partial C_\ell^{ij}}{\partial \theta_\alpha} ,\eq
where $\theta_\alpha$ is the vector of parameters in the theoretical model. All parameters are fully marginalized over.

\section{Parameter estimation results}
\label{Parameter_estimation}

Parameter estimation through Fisher matrix analysis uses a fiducial model which is assumed to be correct. Our `prior knowledge' is included in the calculation through our choice of the assumed model, based on theory and knowledge from previous experiments. In this study we use a fiducial model based on \textit{Planck} 2013 results, which themselves incorporate prior ranges motivated by theory and observation \citep[see][]{Planck-Collaboration:2013aa}. 

\begin{table}
\caption{Predicted $1\sigma$ errors on cosmological parameters using cosmic shear data for two \textit{Euclid} survey configurations, as described in the text.}
\begin{tabular}{llll}
 \hline
Parameter & Fiducial &   \multicolumn{2}{c}{$1\sigma$ errors} \\
&  value &  Requirements &  Goals\\
\hline
$\Omega_m$ & 0.31 & 0.0074 & 0.0053 \\
$\Omega_b$ & 0.048 & 0.0257 & 0.0185 \\
$m_\nu$/$\mr{eV}$ & 0.25 & 0.4074 & 0.2925 \\
$N_\nu$ & 3 & 4.7330 & 3.4393 \\
$h$/$100\,\mr{km}^{-1}\mr{Mpc}^{-1}$ & 0.67 & 0.3277 & 0.2485 \\
$\sigma_8$ & 0.82 & 0.0088 & 0.0062 \\
$n_s$ & 1 & 0.1516 & 0.1129 \\
$\alpha$ & 0 & 0.0491 & 0.0360 \\
$\Omega_\mr{DE}$ & 0.69 & 0.0562 & 0.0384 \\
$w_0$ & -1 & 0.0551 & 0.0382 \\
$w_a$ & 0 & 0.3824 & 0.2559 \\
\hline
FOM &  & 48 & 102 \\
\hline
\end{tabular}
\label{marg}
\end{table}

In this section we present the forecast marginal errors on cosmological parameters using tomographic cosmic shear. 

The predicted marginalised parameter errors on the cosmological parameters in the 11-parameter model are shown in Table \ref{marg}. We show the results for the \textit{Euclid} `requirement' and `goal' survey configurations. The improvement in the constraints, especially for the dark energy equation of state parameters, is due mostly to the increase in the survey area.

The concept of a well-measured parameter depends on the prior. Using a \textit{Euclid} survey with goal parameters, we obtain a posterior uncertainty around $w_0=-1$ and $w_a=0$ of $0.038$ and $0.256$ respectively. Whether this means that we have measured dark energy to be a cosmological constant $\Lambda$ or whether some other model is favoured depends on our model predictions. For instance, $w\geq -1$ is expected for standard minimally coupled scalar fields, while $w<-1$ is allowed in some theories. This dependence of model selection conclusions on the prior range is an important aspect of modern cosmology \citep[see e.g.][]{Kunz:2006aa}.
 
\begin{figure}
\begin{center}
\includegraphics[width=84mm]{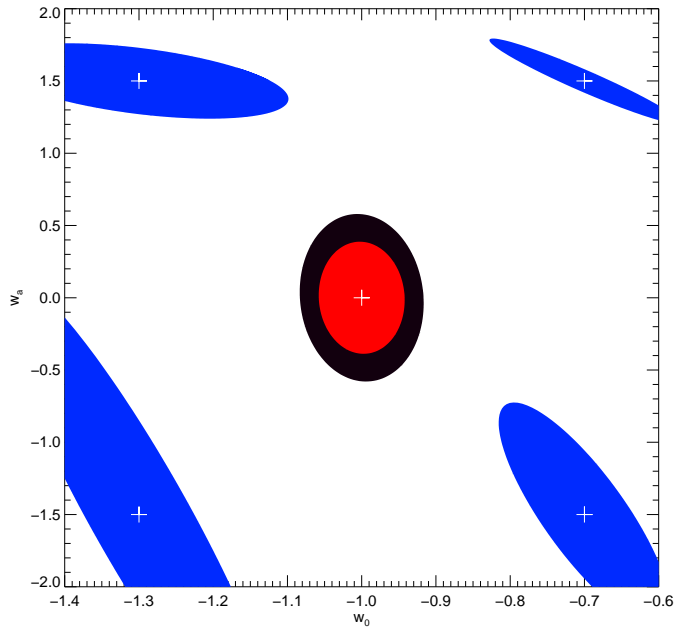}
\end{center}
\caption{The forecast joint $1\sigma$ constraints on the $(w_0,w_a)$ plane from \textit{Euclid}. The blue ellipses show the constraints at the extreme edges of the parameter space we explore. The central ellipses show the results for fiducial values of $w_0=-1$ and $w_a=0$ with the \textit{Euclid} requirement parameters (black) and goal parameters (red). The position of the fiducial dark energy parameter values in each case is shown by the white cross-hairs.}
 \label{ellipses}
\end{figure}

The forecast $1\sigma$ constraints in the $(w_0,w_a)$ plane for particular choices of cosmological model are shown in Fig. \ref{ellipses}. We note that the orientation of the error ellipses shifts as we change the point around which the Fisher matrix is calculated. Around $(w_0=-1,w_a=0)$, the errors on each parameter are only minimally correlated.  

\begin{figure*}
\begin{center}
\includegraphics[width=120mm,angle=270]{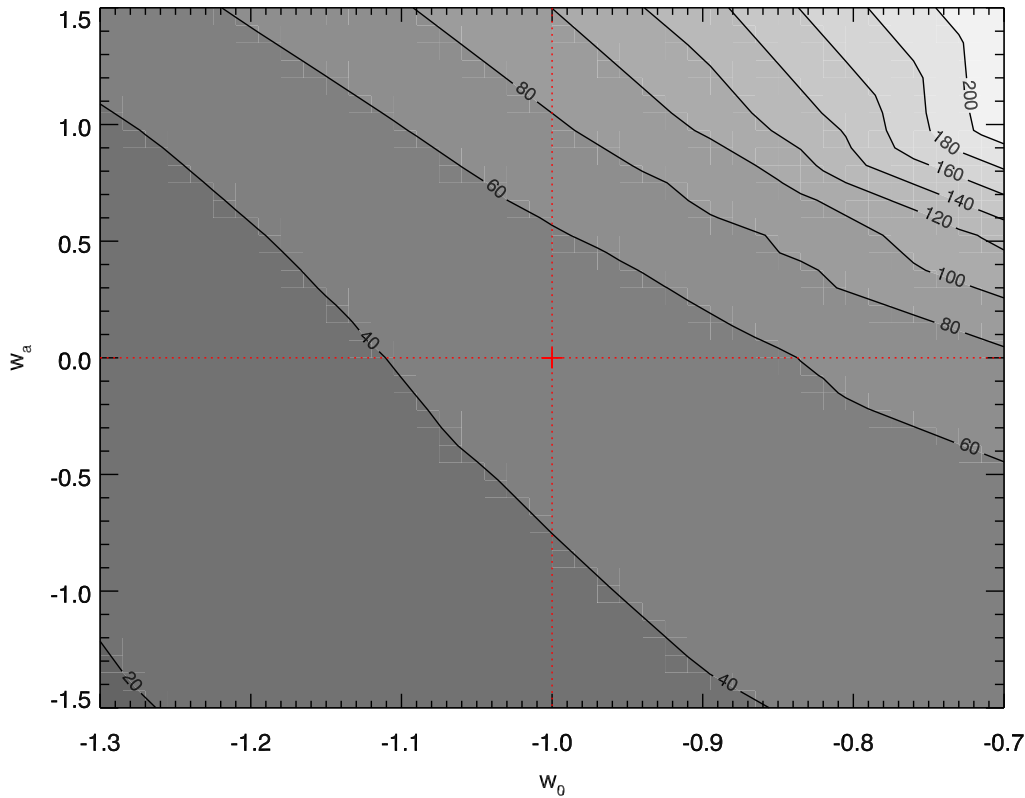}
\end{center}
\caption{The DETF Figure of Merit in the $w_0$-$w_a$ plane, from expected \textit{Euclid} results with the requirement survey configuration. The position of $\Lambda$CDM in parameter space is shown by the red cross-hairs.}
 \label{FOM_2D}
\end{figure*}

Fig. \ref{FOM_2D} shows the predicted Dark Energy Task Force Figure of Merit \citep{DETF} from a \textit{Euclid} survey with requirement configuration in the parameter space spanned by $w_0$ and $w_a$. The Figure of Merit (FOM) quantifies the potential for a survey to constrain dark energy parameters for a dynamical dark energy equation of state parametrized by
\bq w(a)=w_n+(a_n-a)w_a, \eq where $a_n$ corresponds to a pivot redshift at which $w_a$ and $w_n$ are uncorrelated. The FOM is defined as
\bq \mr{FOM}=\frac{1}{\sigma w_n\sigma w_a}. \eq

The FOM is stable over a wide region around $(-1,0)$ in the  $(w_0,w_a)$ plane roughly corresponding to the marginalized posterior distributions for $w_0$ and $w_a$, obtained using combined $\textit{Planck}$ CMB and other data \citep{Planck-Collaboration:2013aa}. Over this region, the marginalized errors for the other parameters in the model do not change significantly, since the second derivative of the log-likelihood in the Fisher matrix is stable as the point around which it is taken is shifted. There is a change in the $(w_0,w_a)$ degeneracy direction, as shown in Fig. \ref{ellipses}, but the joint error remains roughly constant.

It is interesting to note that \citet{Basse:2013aa} obtain qualitatively similar results for the variation of the FOM away from $\Lambda$CDM for a \textit{Euclid}-like survey. They use a different method to calculate the FOM, based on Monte Carlo analysis of mock datasets for cosmic shear power spectrum, galaxy power spectrum, and cluster mass function measurements.

Measuring the FOM alone does not tell us anything about the evidence for the cosmological model. It simply tells us how well a given experiment is able to constrain the parameters of that model. All we can conclude is that if we assume a non-$\Lambda$ fiducial model within some region around $\Lambda$, then the $1\sigma$ confidence limits do not rule out $\Lambda$. The Fisher matrix assumes $\Lambda$CDM (or some other model) is true and simulates the error bounds from it. The error ellipse from parameter estimation is not invariant when changing models. This is clearly seen in Fig. \ref{ellipses}. Bayesian model selection simulates errors from all dark energy models, assesses model confusion, allows us to discriminate against $\Lambda$CDM.This is examined in the next section. 

\section{Bayesian evidence results}
\label{Model_selection}

In the results presented here, the magnitude of the Bayes factor is the ability of a \textit{Euclid}-type experiment to distinguish between one model over another. In other words, this is the evidence provided by the experiment for the fiducial model over a competing model, where the fiducial model contains extra parameters (in this case, dark energy parameters $w_0$ and $w_a$).

One important point is that the prior ranges should be wide enough to fit most of the parameter likelihood. Our choice of prior ranges for $w_0$ and $w_a$ follows that of the \textit{Planck} 2013 calculations \citep{Planck-Collaboration:2013aa}, where the prior ranges are chosen to be larger than the posterior without crossing into regions of parameter space which are unphysical. This means that the prior ranges are also motivated by the structure of our cosmological theory and thus incorporate the concept of `prior knowledge'. We set a prior range of $-3\leqslant w_0 \leqslant -0.3$ and $-2\leqslant w_a \leqslant 2$. The boundary of the decisive region is not significantly affected by changes in the prior range.

\subsection{Single-parameter expected evidence}

In this section we investigate the expected evidence for the parameters $w_0$ and $w_a$ individually.

\begin{figure}
\begin{center}
\includegraphics[width=72mm,angle=270]{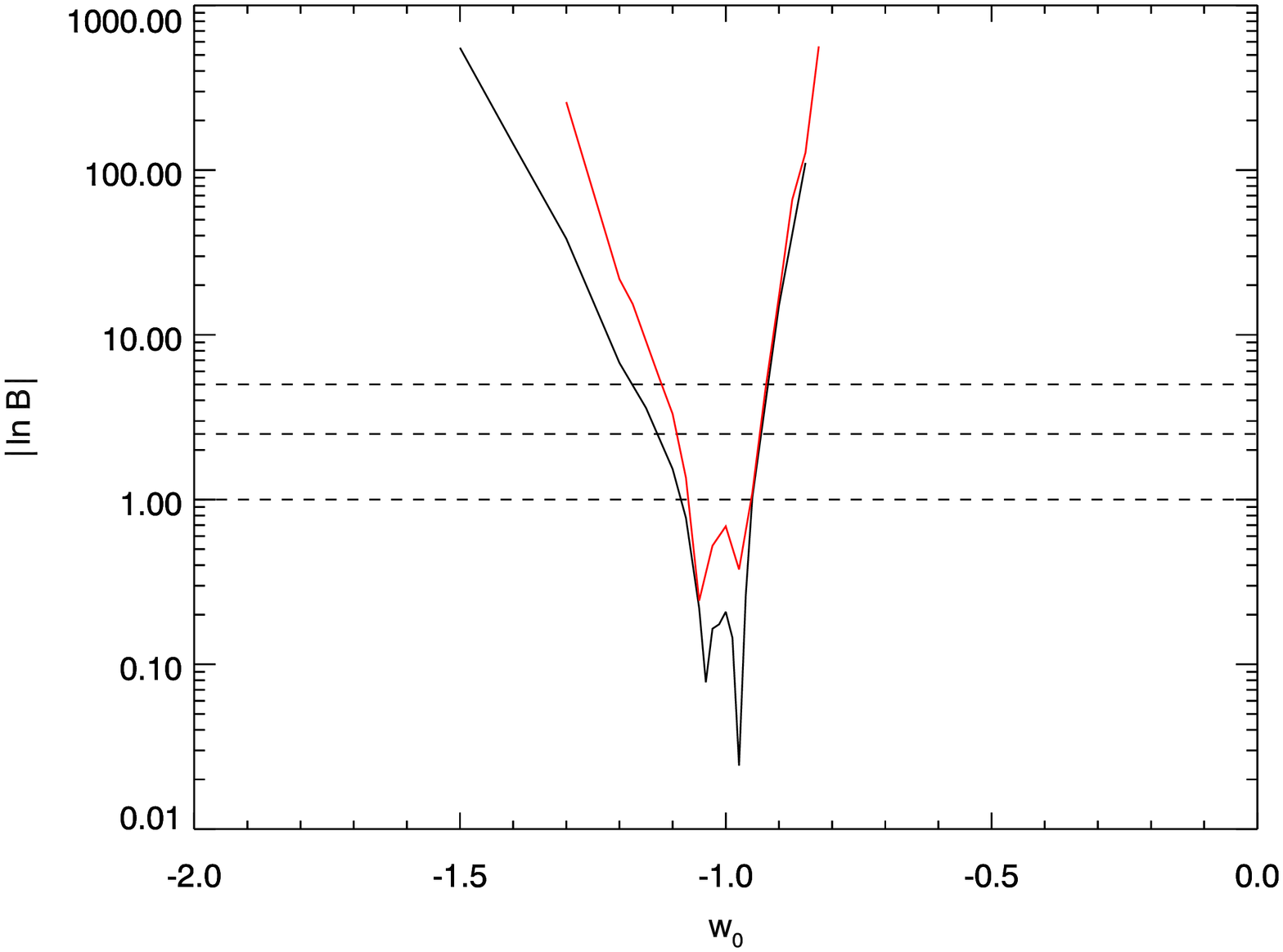}
\includegraphics[width=72mm,angle=270]{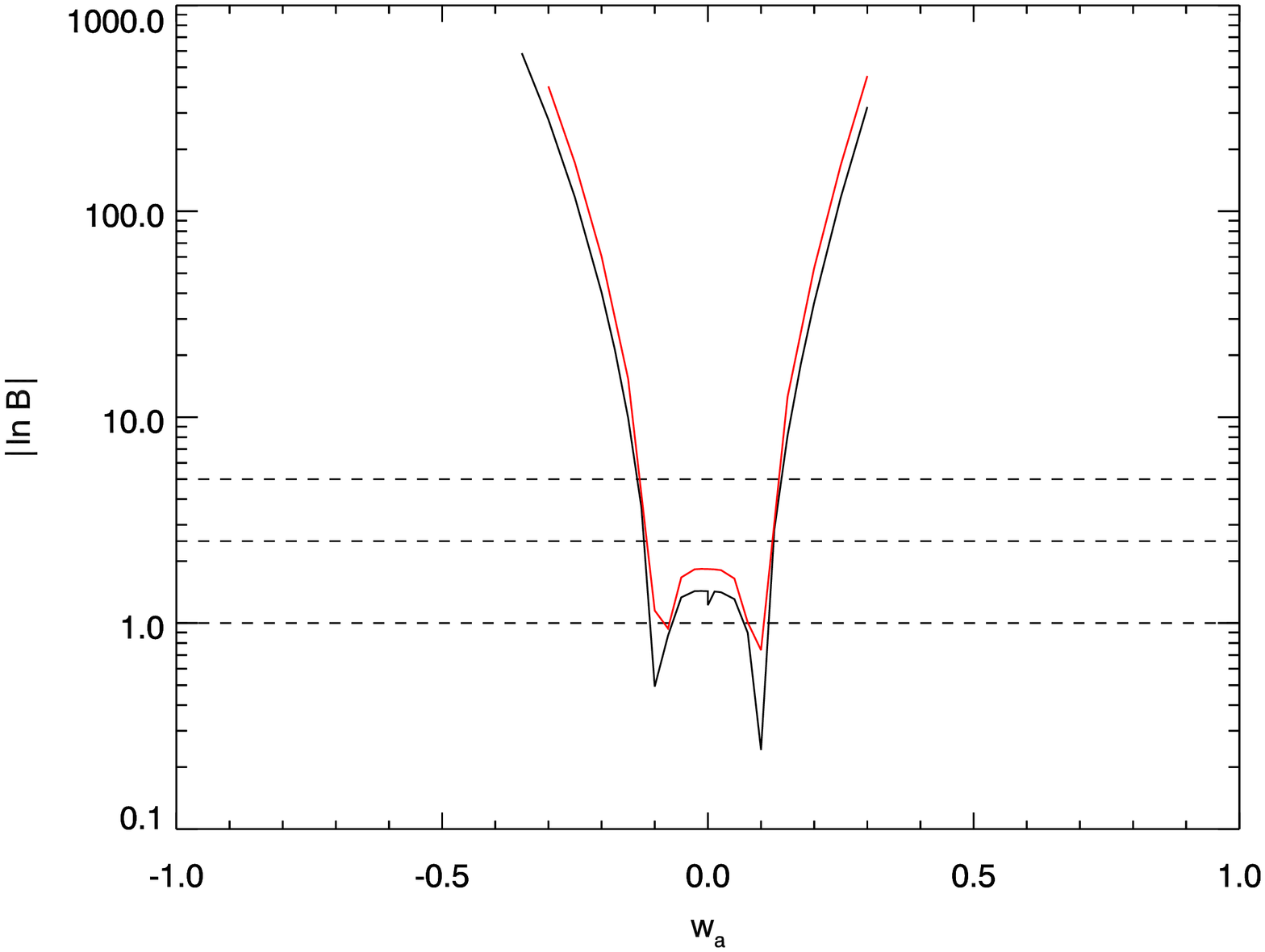}
\end{center}
\caption{The expected value of $| \ln B|$ as a function of the extra parameter in the competing model. In the top panel, we show the evidence for models with constant $w_0$ against $\Lambda$CDM as a function of the value of $w_0$ i.e. hypothesis $H_1$ against $H_0$. In the bottom panel, we show the evidence for quintessence models with nonzero $w_a$ against models with constant $w$ (not necessarily $\Lambda$) as a function of the value of $w_a$ (bottom panel) i.e. hypothesis $H_2$ against $H_1$. The evidence is calculated from forecast \textit{Euclid} data with requirement survey parameters (black) and \textit{Euclid} goal parameters (red). The dotted horizontal lines mark the boundaries between `significant', `strong', and `decisive' evidence according to Jeffreys's terminology. Between the two cusps on each plot, $\Lambda$CDM would be likely to be preferred by the data. }
 \label{Bayes_1D}
\end{figure}

The top panel of Fig. \ref{Bayes_1D} shows the one-dimensional expected evidence for a constant $w_0\neq -1$. In this case, we are testing hypothesis $H_1$ against $H_0$, that is, a 10-parameter model where $w_0$ is constant but non-$\Lambda$ against a 9-parameter $\Lambda$CDM model. 

We find that \textit{Euclid} requirement survey data would decisively discard $\Lambda$CDM for $w_0<-1.176$ or $w_0>-0.921$. Data from the goal survey configuration would narrow the range around $-1$ where the evidence is not decisive to $-1.121\geqslant w_0\geqslant -0.924$. Near $w_0=-1$, the evidence for $\Lambda$CDM begins to accumulate, but the curve stays below $-1$, showing that the evidence is still inconclusive, and neither model is favoured.

The parameter estimation results for the 10-parameter model give us 3$\sigma$ error bounds of $w_0=-1\pm 0.165$ (requirements) and $w_0=-1\pm 0.114$ (goals), equivalent to $99.5$ per cent probability that the true value of $w_0$ falls within these bounds. The ranges are therefore $[-1.165,-0.835]$ (requirements) and $[-1.114,-0.886]$ (goals). This could be mistaken for an example of Jeffreys-Lindley's paradox \citep{Jeffreys:1939,Lindley:1957}; there are values of $w_0$ lying inside the $99$ per cent confidence region around $\Lambda$ for which $\Lambda$CDM would nonetheless be discarded by model selection. In reality, there is no paradox. Here we are calculating evidence \textit{against} $\Lambda$CDM, while the parameter estimation results assume that $\Lambda$CDM is the true model. The Bayes factor is interpreted as a $99.3$ per cent probability that $w_0$ is \textit{not} $\Lambda$ if the data fit models where $w_0$ is outside the range $[-1.176,-0.921]$ (requirements) and $[-1.121,-0.924]$ (goals). Inside these ranges, the experiment is inconclusive. Note that tighter parameter estimation bounds around $-1$ would still be insufficient to rule out alternatives to $\Lambda$CDM if we consider that the prior range for $w_0$ is zero for this model.  This is a good illustration of the need for a multipronged approach to Bayesian inference in cosmology. Parameter estimation provides information which allows us to update our priors, which we can then use for model selection.

Constant $w$ models are not physically well-motivated. If $w\neq -1$ then it is likely to change with time. We therefore investigate the expected evidence for time-varying dark energy equation of state models against constant-$w$ models.

The bottom panel of Fig. \ref{Bayes_1D} shows the one-dimensional expected evidence for $w_a$ using \textit{Euclid} with survey parameters defined by `requirements' (black line) and `goals' (red line). Here we are testing hypothesis $H_2$ against $H_1$ (see Section \ref{Intro}), that is, an 11-parameter model which includes $w_0$ and $w_a$ against a 10-parameter model which includes some constant $w_0$ but not $w_a$. 

We find that \textit{Euclid} data with the requirement survey would decisively favour a $(w_0,w_a)$ model if $w_a$ is outside the range $[-0.133,0.139]$. With the goal survey configuration, this range narrows to $[-0.128,0.134]$. The value of $| \ln B|$ rises above $1$ for $-0.067<w_a<0.068$ (requirements) and $-0.072<w_a<0.075$. Inside this range, the model simplicity criterion (Occam's razor) will significantly favour the simpler model unless the data demand otherwise. The evidence, however, is not decisive, meaning that \textit{Euclid} on its own is not powerful enough to decisively favour a constant-$w$ model over a quintessence model even if $w_a=0$ is the true case. We note that there is no tension between our Bayesian evidence and our parameter estimation results, which give us marginalized $1\sigma$ likelihood of $w_a=0\pm0.38$ and $w_a=0\pm0.26$ for the 11-parameter model with the \textit{Euclid} requirement and goal survey, respectively.

\subsection{Multi-parameter expected evidence}

\begin{figure*}
\begin{center}
\includegraphics[width=140mm,angle=270]{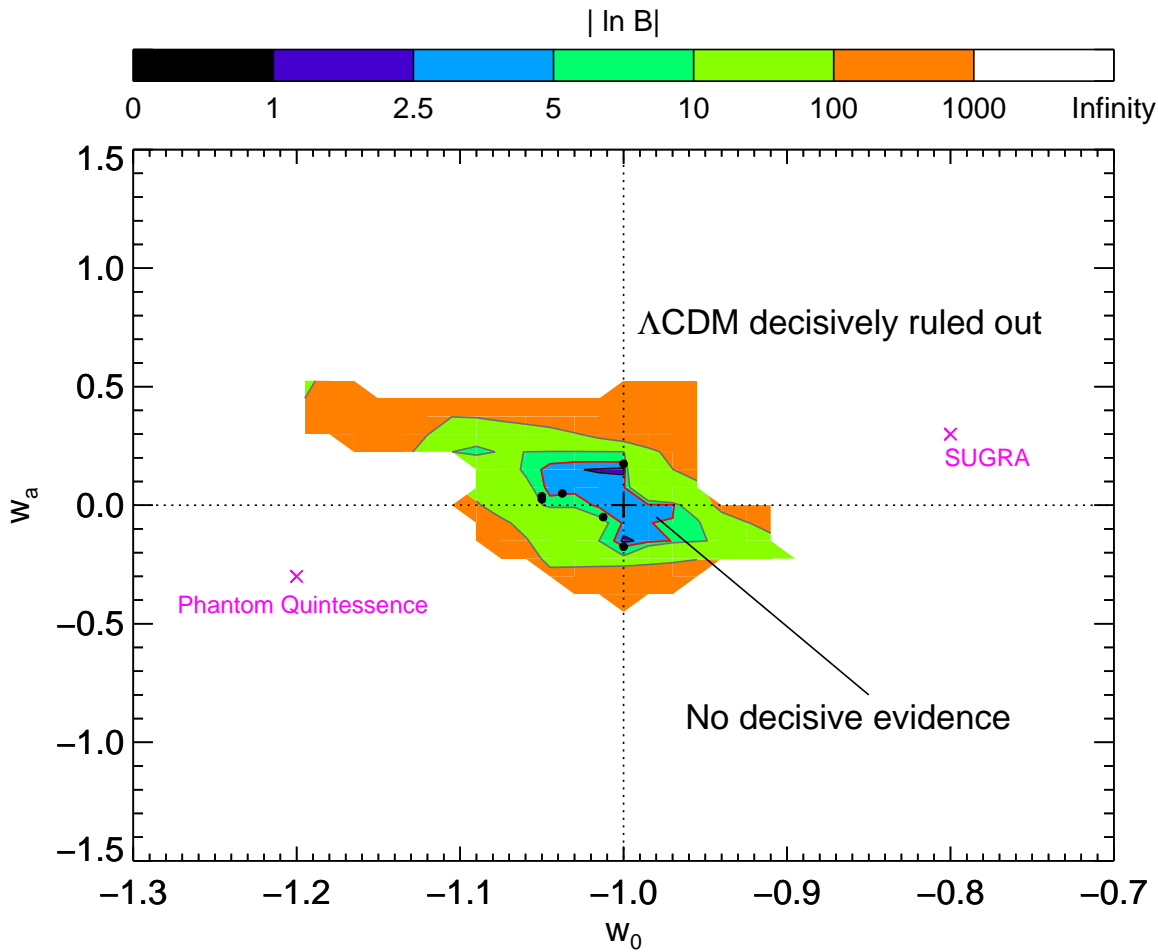}
\end{center}
\caption{The evidence contours for $w$CDM against $\Lambda$CDM, from forecast \textit{Euclid} data with the requirement survey parameters. The position of $\Lambda$CDM in parameter space is shown by the black cross-hairs. The white region indicates the points in parameter space where the numerical value of $| \ln B |$ is close to infinity i.e. where the odds are infinitely against $\Lambda$CDM. Outside the light blue region delineated by a red contour ($|\ln B| =5)$, the evidence is decisively against $\Lambda$CDM. Inside this region, the experiment provides moderate to strong evidence, but the evidence is not decisive. The black points surrounding the $|\ln B| =5)$ contour indicate values of $|\ln B|<1$, i.e. points where the evidence is inconclusive and neither model is favoured. The value of $|\ln B|$ is above unity close to $\Lambda$CDM, indicating positive support for the simpler model. SUGRA and Phantom Quintessence are indicated in magenta. They fall within the region where the evidence is overwhelmingly decisive.} 
 \label{Bayes_2D}
 \end{figure*}

Fig. \ref{Bayes_2D} shows the expected evidence contours for $w_0$ and $w_a$ jointly from the \textit{Euclid} requirements survey design. Here we are testing hypothesis $H_2$ (11-parameter $w$CDM) against $H_0$ (9-parameter $\Lambda$CDM). 
 
We note that the evidence contours are not perfectly symmetric around $\Lambda$CDM and that the Bayes factor values show some isolated minima. At certain points, the value of $|\ln B|$ dips below $1$. These are located close to $\Lambda$ and indicate the points where the evidence in favour of either model is inconclusive. These points are located at the boundary between the parameter space where the data would favour $\Lambda$CDM and the space where the data would favour a dynamical dark energy model. As the models move closer to $(w_0=-1,w_a=0)$, the data will start to favour $\Lambda$CDM because Occam's razor (favouring the simpler model) will now be dominant over the ability of the more complicated model to fit the data with its extra parameters. The figure also shows that evidence accumulates quickly as we move away from the simpler model, which is to be expected -- a \textit{Euclid}-like survey should be able to detect significant deviations from $\Lambda$CDM. This rate of change over parameter space is an indication of the strength of the experiment.

The fiducial survey will be able to distinguish decisively between dynamical dark energy models and $\Lambda$CDM over much of the parameter space examined in our calculations. This space is indicated by the light and dark green, orange and white regions in Fig. \ref{Bayes_2D}. Specifically, the experiment could conclusively distinguish between $\Lambda$CDM and SUGRA \citep{Brax:1999aa} as proposed by \citet{Weller:2002aa} to represent quintessence with $(w_0=-0.8,w_a=0.3)$ or $\Lambda$CDM and Phantom Quintessence \citep{Caldwell:2002aa} with $(w_0=-1.2,w_a=-0.3)$ (see \citet{Novosyadlyj:2013aa} for model selection applied to quintessence and phantom models using other astrophysical probes).

\section{Conclusion}

In this paper we have shown that tomographic weak lensing has the ability to measure the effect of dynamical dark energy on the matter power spectrum, and use this to place constraints on cosmological parameters including dynamical dark energy. We have investigated the expected parameter constraints on dynamical dark energy equation of state parameters from the future \textit{Euclid} weak lensing probe with forecast cosmic shear data, using the Fisher matrix approach.

Using the Bayesian evidence method described in \citet{Heavens:2007}, we have calculated the Savage-Dickey density ratio for nested cosmological models which include cold dark matter, baryons, neutrinos, possible spatial curvature and variations of the primordial scalar spectral index, as well as different parametrizations of the dark energy equation of state.

We have shown that \textit{Euclid} cosmic shear data from the requirement survey would be able to decisively distinguish between a cosmological constant with $w_0=-1$ and a constant but non-$\Lambda$ dark energy equation of state parameter if $w_0<-1.176$ or $w_0> -0.921$.  Our results also show that these data would be able to decisively distinguish between time-constant equation of state models with $w_a=0$ and dynamical dark energy models if $w_a<-0.133$ or $w_a>0.139$.

We find that \textit{Euclid} cosmic shear data with the requirement survey configuration, would be able to provide substantial to strong Bayesian evidence to distinguish dynamical dark energy models from $\Lambda$CDM for models within the space spanned by the $(w_0=-1,w_a=0)$ error ellipse. However, the evidence in this region is still not decisive.

$\Lambda$CDM is well-supported by the forecasts, since the inclusion of extra parameters is not required by Bayesian evidence at the current Concordance Cosmology parameter values. This is shown by the fact that the Bayesian evidence calculations in the parameter space close to $\Lambda$CDM return an undecided result, or a result showing substantial to strong evidence.

This article concludes that a future all-sky weak lensing survey of the $\textit{Euclid}$ type could provide robust constraints on dark energy parameters and distinguish between a wide range of dynamical dark energy models and a cosmological constant. Our results show that if the dark energy equation of state parameter $w$ is really different from $\Lambda$, then this survey is very likely to be able to confirm this, but if dark energy really is $\Lambda$ then \textit{Euclid} tomographic cosmic shear alone will not be decisive.

\section*{Acknowledgements}
The author is supported by the European Space Agency International Research Fellowship.

%\bibliographystyle{aa}
%\bibliography{Debono_Ivan_Arxiv}

%\bsp
\label{lastpage}

\end{document}